\documentstyle[12pt]{article}
\def\Pf{\mbox{Pf}}
\def\det{\mbox{det}}
\def\sign{\mbox{sign}}
\def\ba{\begin{array}}
\def\ea{\end{array}}
\def\bea{\begin{eqnarray}}
\def\eea{\end{eqnarray}}
\def\bea*{\begin{eqnarray*}}
\def\eea*{\end{eqnarray*}}
\def\be{\begin{equation}}
\def\ee{\end{equation}}
\def\si#1{\sigma(#1)}
\def\wsi#1{\widetilde{\sigma}(#1)}
\def\si#1{\sigma(#1)}
\def\wsi#1{\widetilde{\sigma}(#1)}

\begin{document}
\
\vspace{3cm}
\begin{flushright}
Preprint ITP-96-01E\\
hep-th/9601106
\end{flushright}
\vspace{0.5cm}
\renewcommand{\thefootnote}{\arabic{footnote}}

\begin{center}
{{\Large \bf
On Duality  of Two-dimensional Ising Model}}\\
{{\Large \bf
on   Finite Lattice}}\\
\bigskip
{ A.I.~Bugrij
\footnote
{E-mail address:
abugrij@gluk.apc.org }
, V.N.~Shadura}\\
\medskip
 {\it Bogolyubov Institute for Theoretical Physics}\\
 \medskip
 {\it 252143 Kiev-143, Ukraine}
\end{center}
\bigskip
\bigskip
\begin{abstract}
\begin{sloppypar}
It is shown that the partition function of the 2d  Ising model on the dual
finite lattice with periodical boundary conditions is
 expressed through some specific combination of the partition
 functions of the model on the torus with  corresponding boundary
 conditions.
 The generalization of the duality relations for the nonhomogeneous case
is given. These relations
are proved for the weakly nonhomogeneous distribution of the coupling
constants for the finite lattice of arbitrary sizes.
Using the duality relations for the nonhomogeneous Ising model,  we
obtain the duality relations  for the two-point correlation function on the
torus,  the 2d Ising model  with magnetic fields  applied to the boundaries
and  the 2d Ising model with free, fixed and mixed  boundary conditions.

 \end{sloppypar} \end{abstract} \thispagestyle{empty}

\newpage
\pagenumbering{arabic}
\begin{center}
{\Large \bf 1. Introduction}
\end{center}

\medskip
The duality relation for the two-dimensional Ising model was discovered
by Kramers and Wannier [1] in 1941 year. They  established the
correspondence
between the partition function of the model in low-temperature phase and
the partition function in high-temperature phase
\be
(\sinh
2\widetilde{K}))^{-N/2}\widetilde{Z}(\widetilde{K})=
(\sinh
2K)^{-N/2}Z(K)
\ee
$$
\sinh2K\cdot\sinh2\widetilde{K}=1.
$$
Using this self-duality property, in [1] the critical temperature
in the 2d Ising model was determined before  the Onsager`s exact solution [2].

In paper [3]  Kramers-Wannier duality relation (1) was generalized to
the nonhomogeneous case (the coupling constants are arbitrary functions of
lattice site coordinates)
\be
\prod_{\widetilde{r},i} \bigl(\sinh
2\widetilde{K}_i(\widetilde{r})\bigr)^{-1/4}\widetilde{Z}\bigl[
\widetilde{K}_i(\widetilde{r})\bigr]=
\prod_{r,i} \bigl(\sinh {2}K_i(r)\bigr)^{-1/4}Z\bigl[K_i(r)\bigr],
\ee
$$
\sinh 2K_1(r)\cdot\sinh2\widetilde{K}_2(\widetilde{r})=1,
$$
\be
\sinh 2K_2(r)\cdot\sinh2\widetilde{K}_1(\widetilde{r})=1.
\ee
Here $r$, $\widetilde{r}$ and $K_1(r)$, $\widetilde{K}_1(\widetilde{r})$
are  coordinates and coupling constants on the initial and dual
lattices respectively (we will define them in the following section).
Since Kadanoff-Ceva anzats (2) defines connection between functionals but no
functions, this relation is very usefull for analysis of the
thermodymamic phases of the model.  So, for example, this anzats allows to
 correctly define disorder parameter $\mu$, to obtain the duality relation
 connecting correlation functions on the initial and dual lattices, to define
"mixed" correlation functions
$\langle\sigma(r_i)\dots\sigma(r_j),\mu(r_k)\dots\mu(r_l)\rangle$ and e.s.

As was already mentioned in [1,3]  relations (1) and (2)
can not be understood as literal.
So, for example, using method of comparing of high- and
low-temperature expansions for deriving of duality relation (1),
 it is hard to
take into account and to compare the graphs, which include spins on the
boundaries (in particularly, graphs which contain cycles wrapping up the
torus). In fact (1) is correct in the thermodynamic limit (for
the specific free energy).  However for the nonhomogeneous case the procedure
 of thermodynamic limit is rather ambiguous.  In any case it is usefull to have
 exact relations (in contrast to (1) and (2)) connecting the partition
 functions on the initial and dual finite lattices. This is aim of our
 paper.

In the section 2 we introduce  denotions and define the representation of the
partition function in the form of the functional Grassmann integral.
In section 3 we derive exact duality relation  for the homogeneous Ising
model on  the finite lattice with periodical boundary conditions.
 It is shown that the partition function of the model on the dual lattice is
 expressed through specific combination of the partition functions on
 the initial lattice on the torus with the corresponding boundary conditions.
In section 4 we propose the generalization of this relation for
the nonhomogeneous case. Here we prove this relation for the weakly
nonhomogeneous distribution of the coupling constants.
In section 5  the exact duality relation between two-point
order-order and disorder-disorder correlation functions  on the
torus is derived. In section 6 the duality relations
 for  the 2d Ising
model  with   magnetic fields  applied to the boundaries
and  the 2d Ising model with free, fixed and mixed  boundary
conditions are obtained.

\bigskip
\centerline{\Large\bf 2. The model}

\medskip

The partition function of the 2d Ising model on the torus
is
\be
2^NZ[K]=\sum_{[\sigma]}e^{-\beta H[\sigma]},
\ee
\be
-\beta
H[\sigma]=\sum_{r}\sigma(r)\bigl(K_1(r)\nabla_x+K_2(r)\nabla_y\bigr)\sigma(r),
\ee
where $r=(x,y)$    denotes the site coordinate on the square lattice of
size $N=n\times m$,  $x=1,\dots,n$ $y=1,\dots,m$;  $\sigma(r)=\pm1$;
$K_1(r)$ and $K_2(r)$ are coupling constants along horisontal $X$ and
vertical $Y$ axes respectively.
The one-step shift operators  $\nabla_x$, $\nabla_y$
are acting on $\sigma(r)$  in the following way
$$
\nabla_x\sigma(r)=\sigma(r+\widehat{x}),\quad\nabla_y\sigma(r)=\sigma
(r+\widehat{y}),
$$
$$
\nabla_x^T=\nabla_{-x},\quad
\nabla_y^T=\nabla_{-y},
$$
where $\widehat{x}$,  $\widehat{y}$ are unit vectors along
horisontal and vertical axes.
For  periodical boundary conditions along $X$ and $Y$ axes
$$ (\nabla_x^p)^n=1, \quad (\nabla_y^p)^m=1, $$
and for  antiperiodical boundary conditions
$$ (\nabla_x^a)^n=-1, \quad (\nabla_y^a)^m=-1.
$$
Here $p$ and $a$ denote periodical and antiperiodical boundary conditions
respectively.


Since we have the four type of the boundary conditions on the torus
let us assign the corresponding indeces to the partition function:
  $Z^{(\alpha,\beta)}[K]$ and, for example, $$
2^{N}Z^{(a,p)}[K]=\sum_{[\sigma]}\exp\biggl(\sum_{r}
\sigma(r)(K_1(r)\nabla_x^a+K_2(r)\nabla_y^p)\sigma(r)\biggr).
$$
 Later on we will consider
 $Z^{(\alpha,\beta)}[K]$
as the four-component vector
 ${\bf Z}[K]$
 with components
$Z_b[K]$, $b=1,2,3,4$
 \be {\bf
Z}[K]=(Z^{(p,p)},Z^{(p,a)},Z^{(a,p)},Z^{(a,a)}).
\ee
We denote site coordinates, functions and functionals  on the
dual lattice by "tilda" :
$$ \widetilde{r}, \quad
\widetilde{\sigma}(\widetilde{r}), \quad \widetilde{K}_i(\widetilde{r}),
\quad \widetilde{H}[\widetilde{\sigma}], \quad
\widetilde{Z}[\widetilde{K}], \quad \dots $$
The site coordinate on the dual lattice coincides with the coordinate
of the plaquet center on the initial lattice:
$$ \widetilde{r}=r+(\widehat{x}+\widehat{y})/2.
$$
Using these denotions, we can write  the dual partition
function
$$ 2^N\widetilde{Z}[\widetilde{K}]=\sum_{[\widetilde{\sigma}]}
e^{-\tilde{\beta}\widetilde{H}}[\widetilde{\sigma}],
$$
with the dual Hamiltonian
$$
-\tilde{\beta}\widetilde{H}[\widetilde{\sigma}]=\sum_{r}\widetilde{\sigma}(\widetilde{r})
\bigl(\widetilde{K}_1(\widetilde{r})\nabla_{-x}+
\widetilde{K}_2(\widetilde{r})\nabla_{-y}\bigr)
\widetilde{\sigma}(\widetilde{r}).
$$
The coupling constans $K_i(r)$ and $\widetilde{K}_i(\widetilde{r})$
are connected by  duality condition  (3).

It is known that the partition function  of the 2d Ising model can be
represented as the sum of the functional integrals over the lattice real
fermion field [4,5].  For our task it is important that this
 representation can be exactly obtained for the nonhomogeneous model
[6-8] :
\be Z^{(p,p)}[K]={1\over2}\biggl(-Q^{(p,p)}[K]+Q^{(p,a)}[K]
+ Q^{(a,p)}[K]+Q^{(a,a)}[K]\biggr).
\ee
Here $Q^{(\alpha,\beta)}[K]$ is
the Gaussian functional integral over the four-component Grassmann field
:
\be
Q^{(\alpha,\beta)}[K]=\biggl(\prod_{r,i}\cosh K_i(r)\biggr)\int{\cal D}
\psi\exp\bigl(S^{(\alpha,\beta)}[\psi]\bigr), \ee $$ {\cal
D}\psi=\prod_r\prod_{j=1}^4 d\psi_j(r), $$ $\psi_j(r)$ is
Grassmann field :
$$\{\psi_i(r),\psi_j(r')\}=0.
$$
In (8)
the action has the following form
\be S^{(\alpha,\beta)}[\psi]=\sum_{r}\left({\cal
L}^{(0)}\bigl(\psi(r)\bigr)+ {\cal
L}^{(\alpha,\beta)}\bigl(\psi(r)\bigr)\right), \ee $$ {\cal
L}^{(0)}\bigl(\psi(r)\bigr)=\sum_{1\leq i<j}^{4}\psi_i(r)\psi_j(r), $$ $$
{\cal L}^{(\alpha,\beta)}\bigl(\psi(r)\bigr)=
-t_1(r)\psi_3(r)\nabla_x^\alpha\psi_1(r)+t_2(r)
\psi_2(r)\nabla_y^\beta\psi_4(r), $$
$$ t_i(r)\equiv\tanh K_i(r).  $$

Representation (7)  was written for  periodical boundary conditions.
But it is obviously that we can write the similar relations for arbitrary
combinations (periodical and antiperiodical) of the boundary conditions
along $X$ and $Y$ axes.
For example,
 $Z^{(a,p)}$ is distingushed from
$Z^{(p,p)}$ by change of the sign of  coupling constants
 $K_1(r)$ in the last right lattice column :
  $$ Z^{(p,p)}\to
Z^{(a,p)}\quad at \quad K_1(n,y)\to -K_1(n,y),
$$
and therefore
\be
t_1(n,y)\to-t_1(n,y).
\ee
But transformation
 (10) is equvalent to the change  of the boundary conditions for the
 one-step operator in  action (9)
 $$ \nabla^p_x\leftrightarrow
\nabla^a_x $$
Therefore,
 $Z^{(a,p)}$ is distingushed from
$Z^{(p,p)}$ by the arrangement of signs
 $``\pm''$  before terms in
 (7):
 \be
Z^{(a,p)}[K]={1\over2}\biggl(Q^{(p,p)}[K]+Q^{(p,a)}[K] -
Q^{(a,p)}[K]+Q^{(a,a)}[K]\biggr).
\ee
For the rest boundary coditions
($Z^{(p,a)}$, $Z^{(a,a)}$) we have analogous expressions.
Using (6), one can write expressions for the partition functions with
corresponding boundary conditions on the torus in the following compact form:
$$ {\bf
Z}[K]=\widehat{R}{\bf Q}[K].
$$
where we introduce vector
$$
{\bf Q}=\bigl(Q^{(p,p)},Q^{(p,a)},Q^{(a,p)},Q^{(a,a)}\bigr),
$$
and  matrix
$\widehat{R}$ :
\be
\widehat{R}={1\over2}\left(
\ba{rrrr}
-1& 1& 1& 1\\
 1&-1& 1& 1\\
 1& 1&-1& 1\\
 1& 1& 1&-1\ea
 \right),\qquad \widehat{R}^2=1.
\ee

Functional integral (8) is expressed through the Pfaffian of the
$4N\times 4N$-dimensional antysymmetric matrix
 $D$ :
\be
Q^{(\alpha,\beta)}[K]=\biggl(\prod_{r,i}\cosh K_i(r)\biggr)
\Pf(D^{(\alpha,\beta)}),
\ee
where
\be
D^{(\alpha,\beta)}=\left({\ba{cccc}
 0& 1&{1+t_1(r-\widehat{x})\nabla^\alpha_{-x}}& 1\\
-1& 0& 1&{1+t_2(r)\nabla_y^\beta}\\
{-1-t_1(r)\nabla_x^\alpha}&-1& 0& 1\\
-1&{
-1-t_2(r-\widehat{y})\nabla^\beta_{-y}}&-1&\
0\ea}\right)
\ee
and its defines the quadratic form in (9)
$$
S^{(\alpha,\beta)}[\psi]={1\over2}\bigl(\psi,D^{(\alpha,\beta)}\psi\bigr).  $$
Similar expression one can write for the dual lattice :
$$
\widetilde{\bf Z}[\widetilde{K}]=
\widehat{R}\widetilde{\bf Q}[\widetilde{K}],
$$
$$
\widetilde{Q}^{(\alpha,\beta)}[\widetilde{K}]=\biggl(\prod_{\widetilde{r},i}
\cosh\widetilde{K}_i(\widetilde{r}\biggr)\Pf(\widetilde{D}^{(\alpha,\beta)}),
$$
where matrix $\widetilde{D}$ has the following form
\be
\widetilde{D}^{(\alpha,\beta)}=\left(\ba{cccc}
\ 0&\ 1&{
1+\widetilde{t}_1(\widetilde{r})\nabla^\alpha_{-x}}&\ 1\\
-1&\
0&\ 1&{
1+\widetilde{t}_2(\widetilde{r}-\widehat{y})\nabla^\beta_{y}}\\
{
-1-\widetilde{t}_1(\widetilde{r}+\widehat{x})\nabla^\alpha_x}&-1&\
0&\ 1\\
-1&{
-1-\widetilde{t}_2(\widetilde{r})\nabla^\beta_{-y}}&-1&\
0\ea\right) \ee
$$
\widetilde{t}_i(\widetilde{r})\equiv\tanh
\widetilde{K}_i(\widetilde{r})
$$

In conclusion of this section we note that for the finite lattice
the Pfaffian of matrix
 (13) or (15) is the finite power polinom of
  $t_i(r)$ or
$\widetilde{t}_i (\widetilde{r})$ .
It is not hard to calculate
 $\Pf(D)$ in the high-temperature limit :
\be
\Pf(D)=1, \quad  \mbox{when}\quad t_i(r)=0,
\ee
and  $ \Pf(\widetilde{D})$  in the low-temperature limit:
\be
\Pf(\widetilde{D})=1,   \quad  \mbox{when}\quad
\widetilde{t}_i(\widetilde{r})=0.
\ee
\newpage
\bigskip
\centerline{\Large\bf 3. The homogeneous case}

\medskip

In general case for arbitrary sizes
 $m$, $n$  and distributions of the coupling constants
 $K_i(r)$  we can not calculate neither the Pfaffian nor the  determinant of
matrix $D$ .  But in the homogeneous case, when $$ K_i(r)=K_i=const $$
and the matrix $D$ becomes the translation-invariant matrix the
determinant of $D$ is calculated by means of the Fourier transformation:
$$ \det(D)=\prod_{\bf p}\bigl[(1+t_1^2)(1+t_2^2)-2t_1(1-t_2^2)\cos
p_x-2t_2(1-t_1^2)\cos p_y\bigr],
$$
where $t_i\equiv\tanh K_i$
and the momentum components  $p_x, p_y$  (${\bf p}=(p_x,p_y)$)
take the integer and half-integer values (in units of $2\pi/n$ and
$2\pi/m$) for  periodical and antiperiodical boundary
conditions respectively.
Using relation
\be
\bigl(\Pf(D)\bigr)^2=\det(D),
\ee
we obtain  from (14), (18)
\be
Q^2(K_1,K_2)=\prod_{\bf p}
(c_1c_2-s_1\cos p_x-s_2\cos p_y),
\ee
where
$$
c_i\equiv\cosh 2K_i,\quad s_i\equiv\sinh 2K_i.
$$

In similar way one gets for the dual lattice:
\be
\widetilde{Q}^2(\widetilde{K}_1,\widetilde{K}_2)=\prod_{\bf p}
(\widetilde{c}_1\widetilde{c}_2-\widetilde{s}_1\cos
p_x-\widetilde{s}_2\cos p_y).
\ee
Here
$$
\widetilde{c}_i\equiv\cosh 2\widetilde{K}_i,\quad
\widetilde{s}_i\equiv\sinh 2\widetilde{K}_i.
$$
Using (19) and (20), it is not hard to check the following equality
\be
(s_1s_2)^{-N/2}Q^2(K_1,K_2)=(\tilde
{s}_1\tilde {s}_2)^{-N/2}\widetilde{Q}^2(\widetilde{K}_1,
\widetilde{K}_2), \ee where $s_i$ and ${\tilde s}_i $  satisfy  relations
$$ s_1\cdot\widetilde{s}_2=1,\quad s_2\cdot\widetilde{s}_1 =1.
$$
Note
that (21) one can consider as the duality relation for the square of the
functional integrals which appear in representation  (7)  for
the partition function.
Extracting the square root  from the both parts of (21), we obtain
\be
(s_1s_2)^{-N/4}Q^{(\alpha,\beta)}(K_1,K_2)=\pm(\widetilde{s}_1
\widetilde{s}_2)^{-N/4}\widetilde{Q}^{(\alpha,\beta)}(\widetilde{K}_1,
\widetilde{K}_2).
\ee
Show that sign ``$-$'' in (22)  appears only for function
 $Q^{(p,p)}$, but for the rest components of vectors
$\bf Q$ and $\widetilde{\bf Q}$ we have the sign ``$+$'' :
\be
(\widetilde{s}_1\widetilde{s}_2)^{-N/4}
\widetilde{\bf Q}(\widetilde{K}_1,\widetilde{K}_2)=
(s_1s_2)^{-N/4}
\widehat{g}{\bf Q}(K_1,K_2),
\ee
where  signature matrix $\widehat{g}$ has form
\[
\widehat{g}=\left(\ba{cccc}
-1&\ 0&\ 0&\ 0\\
\ 0&\ 1&\ 0&\ 0\\
\ 0&\ 0&\ 1&\ 0\\
\ 0&\ 0&\ 0&\ 1\ea\right).
\]

Since $\Pf(D)$ is the polinom over
$t_i$, then from (19), (20) it follows that $\det(D)$  is square of this
polinom.
Really in product (19) the multipliers appear by pairs according to the
momentum components
$\pm p_x$ and
$\pm p_y$.
The exclusion is multipliers corresponding to the values
$p_x=p_y=\pi$ and $p_x=p_y=0$.
Let us denote them by $q_\pi$ and
$q_0$:
$$
q_\pi=c_1c_2+s_1+s_2=(1+t_1+t_2-t_1t_2)^2\cosh^2K_1\cdot\cosh^2K_2
$$
$$
q_0=c_1c_2-s_1-s_2=(1-t_1-t_2-t_1t_2)^2\cosh^2K_1\cdot\cosh^2K_2=
$$
$$
=(1-s_1s_2)^2/q_\pi.
$$
Hence it is clear that
in  functions
$Q^{(\alpha,\beta)}$
(except of $Q^{(p,p)}$)
all multipliers have the constant sign in the following range of
parameter values :
$$
s_1\geq0,\quad s_2\geq0,
$$
and, therefore, these functions do not change the sign.
On the other hand
 at crossing through
 critical line  $s_1s_2=1$
 function $Q^{(p,p)}$
which contains  multiplier
 $(q_0)^{1/2}\sim(1-s_1s_2)$ changes the sign .
 Similar results we obtain for  dual functions
$\widetilde{Q}^{(\alpha,\beta)}$.

Therefore, taking into account the high-temperature and low-temperature
limits (16) and (17) for the  Pfaffian, one gets
$$
\sign\bigl(Q^{(p,p)}(K_1,K_2)\bigr)=\sign(1-s_1s_2) $$ $$
\sign\bigl(\widetilde{Q}^{(p,p)}(\widetilde{K}_1,\widetilde{K}_2)\bigr)=
\sign(1-\widetilde{s}_1\widetilde{s}_2)=
$$
$$
=-\sign\bigl((Q^{(p,p)}(K_1,K_2)\bigr).
$$
This relation proves (23).

Multiplying the right and left parts of
(23) on the matrix $\widehat{R}$ (12),
we obtain  the duality relation for the partition functions:
\be
(\widetilde{s}_1\widetilde{s}_2)^{-N/4}
\widetilde{\bf Z}(\widetilde{K}_1, \widetilde{K}_2)=
(s_1s_2)^{-N/4}
\widehat{T}{\bf Z}(K_1,K_2),
\ee
$$
\widehat{T}=\widehat{R}\widehat{g}\widehat{R},\quad\widehat{T}^2=1, $$
\be \widehat{T}={1\over 2}\left(\ba{rrrr} \ 1&\ 1&\ 1&\ 1\\
\ 1&\ 1&-1&-1\\ \ 1&-1&\ 1&-1\\ \ 1&-1&-1&\ 1\ea\right).
\ee
From (24) it follows that difference between the right and left parts of
Kramers-Wannier duality relation (1)
$$
(s_1s_2)^{-N/4}Z^{(p,p)}(K_1,K_2)-(\widetilde{s}_1\widetilde{s}_2)
^{-N/4}\widetilde{Z}^{(p,p)}(\widetilde{K}_1,\widetilde{K}_2)=
$$
\be
=(\widetilde{s}_1\widetilde{s}_2)^{-N/4}\widetilde{Q}^{(p,p)}(\widetilde{K}_1,
\widetilde{K}_2)=-(s_1s_2)^{-N/4}Q^{(p,p)}(K_1,K_2)
\ee
is equal to zero only on  critical line
$$
\sinh 2K_1\cdot\sinh 2K_2=1.
$$
Moreover outside this line
the right and left parts of  (26)  are compared
among themselves
at increasing the lattice size
($m,n\to \infty$).
Therefore duality relation (1) is correct only in the thermodynamic limit:
$$
\lim_{m,n\to\infty} \left({1\over
N}\ln\biggl({\widetilde{Z}(\widetilde{K}_1,
\widetilde{K}_2)\over(\widetilde{s}_1\widetilde{s}_2)^{N/4}}\biggr)\right)=
\lim_{m,n\to\infty}
\left({1\over
N}\ln\biggl({Z(K_1,K_2)\over(s_1s_2)^{N/4}}\biggr)\right).
$$
Since for the finite lattice partition functions $Z^{(\alpha,\beta)}$
are analytic functions of $K_i$  (the finite power polinom over $e^{\pm K_i}$)
the nonanalytic  dependence  in the right and left parts of relation (24)
can appear only from multipliers $(s_1s_2)^{-N/4}$ .
Therefore, taking into account the circuit rules of branching points
corresponding to
$\sinh2K_1=0$ and
$\sinh2K_2=0$,
duality relation (24) can be continued to the rest ranges of values of
the coupling constants:
$K_1\geq0$, $K_2<0$; $K_1<0$, $K_2\geq0$; $K_1<0$, $K_2<0$.

\newpage
\bigskip
\centerline{\Large\bf 4. The nonhomogeneous case}

\medskip

In previous section it was shown that the Kramers-Wannier duality
relation (1) (having rather symbolic meaning) can be modificated to the
exact equality.
The Kadanoff-Ceva  anzats (2) also was proved in [3] for special
(nonphysical) boundary conditions.
Note that the thermodynamic limit for the nonhomogeneous Ising model
can be defined for specific class of the functions determining
distribution of the coupling constants  $K_i(r)$.
For illustration let us consider, for example,  the sequence of functions
 $\{K^{(N)}_i(r)\}$ defined in  the following way:  the coupling
constants are equal zero on the boundary $\Gamma$ of the finite-size clasters:
$K_i(r\in\Gamma)=0$.
In this case at $N\to \infty$  we obtain
the increasing numbers of the non-interacting finite-size clasters.
Then this is  rather similar on the self-averaging procedure for disodered
systems than on the usual thermodynamic limit.
It means that the duality relation for the nonhomogeneous system must be
 formulated for the finite lattice.

The covariant notation of  exact duality relation (24) for the homogeneous
model suggests the obvious recipt for the generalization to the
nonhomogeneous case.
 For finite lattice on the torus  Kadanoff-Ceva anzats (2) must
 be modified  by the following way:
\be
\prod_{\widetilde{r},i}\bigl(\sinh2\widetilde{K}_i(\widetilde{r})
\bigr)^{-1/4}{\widetilde{\bf Z}}[\widetilde{K}]=
\prod_{r,i}\bigl(\sinh2K_i(r)\bigr)^{-1/4}\widehat{T}{\bf Z}[K].
\ee
Unfortunely we can not prove this relation for arbitrary lattice size and
distribution of the coupling constants.
But we checked  duality relation (27) for  lattices of small size
by direct calculation on the computer.
Moreover the duality relation can be proved for the weakly-nonhomogeneous
case:  $$ K_i(r)=K_i+\delta K_i(r),\quad K_i=const,\quad\delta K_i(r)\ll
1, $$
when $n$ and $m$ are arbitrary and finite.

For the first order over
 $\delta K_i(r)$ we have
$$
\prod_{r,i}\biggl(\sinh\bigl(2K_i+\delta
K_i(r)\bigr)\biggr)^{-1/4}=
$$
\be
\mbox{}=(s_1s_2)^{-N/4}\biggl[1-{c_1\over 2s_1}\sum_{r}\delta K_1(r)-
{c_2\over 2s_2}\sum_{r}\delta K_2(r)\biggr],
\ee
$$
{\bf Z}[K]={\bf Z}(K_1,K_2)\biggl[1+\sum_{r}\bigl(
\langle\sigma(r)
\sigma(r+\widehat{x})
\rangle\delta K_1(r)+
$$
\be
\mbox{}+
\langle\sigma(r)
\sigma(r+\widehat{y})
\rangle\delta K_1(r)\bigr)\biggr].
\ee
Note that at different $K_1$ and $K_2$ and (or) $n\neq m$ correlation
functions $\langle\sigma(r) \sigma(r+\widehat{x}) \rangle$ and
$\langle\sigma(r)
\sigma(r+\widehat{y})
\rangle$
do not equal among themselves and due to translation invariance
do not depend from $r$ :
$$
{\bf Z}(K_1,K_2)
\langle\sigma(r) \sigma(r+\widehat{x})
\rangle={1\over N}\widehat{R}{\partial {\bf
Q}(K_1,K_2)\over\partial K_1},
$$
\be
{\bf Z}(K_1,K_2)
\langle\sigma(r) \sigma(r+\widehat{y}) \rangle={1\over
N}\widehat{R}{\partial {\bf Q} (K_1,K_2)\over\partial K_2}.
\ee
Taking into account (28)-(30),  in the first order over
$\delta K_i(r)$ and $\delta \widetilde{K}_i(\widetilde{r})$
we obtain for duality relation
(27):
$$ (s_1s_2)^{-1/4}\left[{\bf Z}(K_1,K_2)
+\widehat{R}\biggl({1\over N}{\partial {\bf Q}\over\partial
K_1}-{c_1\over 2s_1}{\bf Q}\biggr) \sum_{r}\delta K_1(r)+\right.  $$ $$
+\widehat{R}\biggl({1\over N}{\partial
{\bf Q}\over\partial K_2}-{c_2\over 2s_2}{\bf Q}\biggr)
\left.\sum_{r}\delta K_2(r)\right]=
$$
$$
=(\widetilde{s}_1\widetilde{s}_2)^{-1/4}\left[\widetilde{\bf Z}
(\widetilde{K}_1,\widetilde{K}_2)+\widehat{R}\biggl({1\over
N}{\partial \widetilde{\bf Q}\over\partial K_1}-
{\widetilde{c}_1\over2\widetilde{s}_1}{\bf Q}\biggr)\sum_{r}\delta
\widetilde{K}_1(\widetilde{r})+\right.
$$
\be
+\widehat{R}\biggl({1\over N}{\partial
\widetilde{\bf Q}\over\partial K_2}-{\widetilde{c}_2\over
2\widetilde{s}_2}{\widetilde{\bf Q}}\biggr) \left.\sum_{r}\delta
\widetilde{K}_2(\widetilde{r})\right].
\ee
For the order zero terms  this equality is satisfied according to
homogeneous duality relation (24). Show that it is satisfied
for the linear terms over $\delta K$ and $\delta \widetilde{K}$ terms.
From (3) and (23) it follows:
$$
\delta
K_1(r)=-{1\over\widetilde{s}_2}\delta\widetilde{K}_2(\widetilde{r}),\quad
\delta
K_2(r)=-{1\over\widetilde{s}_1}\delta\widetilde{K}_1(\widetilde{r}),
$$
\be
{\partial\over\partial
K_1}=-\widetilde{s}_2{\partial\over\partial\widetilde{K}_2},\quad
{\partial\over\partial
K_2}=-\widetilde{s}_1{\partial\over\partial\widetilde{K}_1},
\ee
\be
{\bf Q}(K_1,K_2)=(\widetilde{s}_1
\widetilde{s}_2)^{-N/2}\widehat{g}\widetilde{\bf Q}
(\widetilde{K}_1,\widetilde{K}_2).
\ee

Substituting (32) and (33) in the left part of  (31) and collecting
similar terms, one gets
 $$ (\widetilde{s}_1
\widetilde{s}_2)^{-N/4}\left[\widehat{T}{\widetilde{\bf Z}}
(\widetilde{K}_1,\widetilde{K}_2)+
\widehat{R}\widehat{g}\biggl({1\over
N}{\partial{\widetilde{\bf Q}}\over\partial\widetilde{K}_2}-
{\widetilde{c}_2\over2\widetilde{s}_2}{\widetilde{\bf Q}}\biggr)
\sum_{\widetilde{r}}\delta\widetilde{K}_2(\widetilde{r})+\right.
$$
\be
+
\left.\widehat{R}\widehat{g}\biggl({1\over
N}{\partial{\widetilde{\bf Q}}\over\partial\widetilde{K}_1}-
{\widetilde{c}_1\over2\widetilde{s}_1}{\widetilde{\bf Q}}\biggr)
\sum_{\widetilde{r}}\delta\widetilde{K}_1(\widetilde{r})\right].
\ee
Using (25), (from which follows
$
\widehat{R}\widehat{g}=\widehat{T}\widehat{R}
$)
it is not hard to show that
 (34)  coincides with the right part of (31).
 This proves duality relation (27) in the weakly-nonhomogeneous case for
 arbitrary $m$ and  $n$.

In conclusion of this section note that for analysis of the duality
relation  for  correlation functions it is convenient to use the other
normalization in (27). Using relations
$$
{\cosh^22K_1(r)\over\sinh2K_1(r)}=
{\cosh^22\widetilde{K}_2(\widetilde{r})\over\sinh2\widetilde{K}_2
(\widetilde{r})},
$$
$$
{\cosh^22K_2(r)\over\sinh2K_2(r)}=
{\cosh^22\widetilde{K}_1(\widetilde{r})\over\sinh2\widetilde{K}_1
(\widetilde{r})},
$$
which follow from (3), and introducing denotions according to [3]
 $$ {\bf
Y}[K]=\prod_{r,i}\bigl(\cosh2K_i(r)\bigr)^{-1/2}{\bf Z}[K], $$ \be
{\widetilde{\bf Y}}[\widetilde{K}]=\prod_{\widetilde{r},i}\bigl(\cosh2
\widetilde{K}_i(\widetilde{r})\bigr)^{-1/2}{\widetilde{\bf Z}}
[\widetilde{K}],
\ee
we obtain instead of (27):
\be
{\bf Y}[K]=\widehat{T}{\widetilde{\bf
Y}}[\widetilde{K}].
\ee

\bigskip
\centerline{\Large\bf 5. Duality relation for  correlation function}

\medskip
The duality relation for the nonhomogeneous Ising model is usefull for study
the correlation function properties.
For this aim it is convenient to use the magnetic dislocation
representation for correlation functions [3]. This representation is
based on the obvious equality
$$
e^{(K+i\pi/2)}\sigma_1\sigma_2=i\sigma_1\sigma_2e^{K\sigma_1\sigma_2}.
$$
Taking into account that
$$
\sigma_1\sigma_n=(\sigma_1\sigma_2)(\sigma_2\sigma_3)
\dots(\sigma_{n-1}\sigma_n),
$$
one can write
$$
\sum_{[\sigma]}e^{-\beta
H[\sigma]}\sigma(r)\sigma(r')=i^{-\gamma}\sum_{[\sigma]}
e^{-\beta H'[\sigma]},
$$
where  Hamiltonian  $\beta H'[\sigma]$ of the Ising model with
defect differs from  $\beta H[\sigma]$
by the change of the coupling constants $K$  on  $K'=K+i\pi/2$
on the defect line $\Gamma_\sigma$ [3]  which connect
sites $r$ and $r'$:
\bea*
K_i(r)&=&K,\\
K'_i(r)&=&\left\{\ba{ll} K+i\pi/2&\mbox{on the links belonging to path
$\Gamma_\sigma$}\\
K&\mbox{on the rest links,}\ea\right. \eea*
$\gamma$ is the length of this path ( the number of
the "spoilt" bonds).
Then, using functionals (35), we obtain  representation for the two-point
correlation function  (we omit boundary condition indeces):
\be
G_\sigma(r,r')\equiv\langle\sigma(r)\sigma(r')\rangle=Y[K']/Y[K].
\ee

In the work [3] it was introduced the  correlation function
of the disorder parameter
$\mu(\widetilde{r})$.
This variable characterizes the degree of disorder near the point
$\widetilde{r}$ on the initial lattice and its one can consider as the
result
of the duality transformation for  Ising spin
$\sigma(r)$.
Correlation function
$\langle\mu(\widetilde{r}),\mu(\widetilde{r'})\rangle$
is determined  by means of  magnetic dislocation
$\Gamma_\mu$:
$$ G_\mu(\widetilde{r},\widetilde{r}')\equiv
\langle\mu(\widetilde{r})\mu(\widetilde{r}')\rangle=Y[K'']/Y[K],
$$
where
\[
K''_i =\left\{\ba{rl} -K&\mbox{on the links intersecting of path
$\Gamma_\mu$}\\
K&\mbox{on the rest linkss}.\ea\right.\]

The duality relation for correlation functions
[3],
\be
\langle\widetilde{\mu}(r)\widetilde{\mu}(r')\rangle=
\langle\sigma(r)\sigma(r')\rangle
\ee
follows from (2)
and the transformation of magnetic dislocation $\Gamma_\sigma$ on the
initial lattice to magnetic dislocation
$\widetilde{\Gamma}_\mu$  on the dual lattice by means of mapping [3]
\bea*
&&K_1(r)+i\pi/2\to \widetilde{K}_2(\widetilde{r})\cdot e^{-i\pi},\\
&&K_2(r)+i\pi/2\to \widetilde{K}_1(\widetilde{r})\cdot e^{-i\pi},
\eea*
which follows from  (3).
Since duality relation
 (27)
 for the finite lattice on the torus differs from Kadanoff-Ceva
anzats (2) the duality relation for correlation functions on the torus
has more complicate form.
For example, using (36), we obtain for the dual lattice with  periodical
boundary conditions
$$
\widetilde{G}_{\widetilde{\mu}}^{(p,p)}(r,r')=\widetilde{Y}^{(p,p)}
[\widetilde{K}'']/\widetilde{Y}^{(p,p)}[\widetilde{K}]=(\widehat{T}{\bf Y}[K'])^
{(p,p)}/(\widehat{T}{\bf Y}[K])^{(p,p)}=
$$
$$
\mbox{}=
\bigl[Z^{(p,p)}G_\sigma^{(p,p)}(r,r')+Z^{(p,a)}G_\sigma^{(p,a)}(r,r')+
Z^{(a,p)}G_\sigma^{(a,p)}(r,r')+
$$
\be
\mbox{}+
Z^{(a,a)}G_\sigma^{(a,a)}(r,r')\bigr]\big/
\bigl[Z^{(p,p)}+Z^{(p,a)}+Z^{(a,p)}+Z^{(a,a)}\bigr].
\ee
It is not hard to show that
 (39) pass to (38) only under the following condition:
 the correlation length is smaller of  the lattice sizes
 (this is happen out the scaling domain and at the large $m$ and $n$).
Note that duality relation
 (39)   coincides with the relation
 for correlation functions on the torus in the
critical point  obtained
in the paper [9] by means of the quantun conformal field theory
methods [10].

\newpage
\bigskip
\centerline{\Large\bf 6. Duality and  boundary conditions}

\medskip

The duality relation for the nonhomogeneous Ising model on the torus allows to
obtain duality relations for the 2d Ising model with magnetic fields applied to
the boundaries and for the 2d Ising model with free, fixed and mixed boundary
conditions.

In order to get these relations let us consider the Ising model on the torus
with defect which is defined by  the following distribution of the coupling
constants in Hamiltonian (5): $K_1(r) =K_2(r)=K$ on all links of
the lattice with the exeption of the following cases -- $K_1(n-1,y)=h_1$,
$K_1(n,y)=h_2$, $K_2(n,y)=h$, where $h_1=\beta H_1$, $h_2=\beta H_2$,
$h=\beta H$, $y=1,\dots,m$.
Using (3), this defect one  can define  on the dual lattice by the following
way:
$\widetilde{K}_1(\widetilde{r})=\widetilde{K}_2(\widetilde{r})=\widetilde{K}$
on all links
with the
exeption of the following cases --
$\widetilde{K}_1(n-1,\widetilde{y})=\widetilde{h}_1$,
$\widetilde{K}_1(n, \widetilde{y})=\widetilde{h}_2$,
$\widetilde{K}_2(n,y)=\widetilde{h}$, where
$\widetilde{h}_1=\beta \widetilde{H}_1$, $\widetilde{h}_2=\beta
\widetilde{H}_2$, $\widetilde{h}=\beta \widetilde{H}$,
$\widetilde{y}=1,\dots,m$ and coupling constants $h$, $h_1$,  $h_2$
and $\widetilde{h}$,   $\widetilde{h}_1$
  $\widetilde{h}_2$  are connected by relations:
\be
\sinh 2h\cdot\sinh2\widetilde{h}=1,\quad
\sinh 2h_1\cdot\sinh2\widetilde{h}_1=1,\quad
\sinh 2h_2\cdot\sinh2\widetilde{h}_2=1.
\ee
Taking limit $h\to\infty$ in partition function (2)
(for simplicity we consider the ferromagnetic model),
it is not hard to obtain
the partition function of the 2d Ising model with  magnetic fields $H_1$ and
$H_2$ aplied to the boundaries:
$$
2 Z^{p}(K,h_1, h_2)=\lim_{h\to\infty} (\cosh h)^{-m}Z^{(\alpha,p)}
(K, h_1, h_2, h)=
$$
\be
2\sum_{[\sigma]}\exp(K\sum_{r,i}\sigma(r)\sigma(r+i)
+h_1\sum_{y=1}^{m}\si{n-1,y}+h_2\sum_{y=1}^{m}\si{1,y}),
 \ee
where in order to sum over spin variables
$\{\si{n,y}\}$ we used the following equality:
\be \lim_{h\to\infty} (\cosh
h)^{-m}\prod_{y=1}^{m} \exp({h\si{n,y}\si{n,y+1}})=
\prod_{y=1}^{m} \delta(\si{n,y},\si{n,y+1}).
\ee
Here on  the right-hand side  the product of Kronecker`s $\delta$-functions
is written. From $\{\si{n,y}\}$  this product  selects
the two spin configurations: all spins are directed up or down.

Note, that in (41)  partition function $Z^{p}(h_1, h_2)$ is obtained by
the limiting procedure with corresponding normalization which removes
 infinite constant. In other cases  in consequence of the conflict between
the product of Kronecker`s $\delta$-functions and the boundary conditions
 we can get the zero after taking of the limit, for example,
\be
\lim_{h\to\infty} (2\cosh h)^{-m}Z^{(\alpha,a)}
(K, h_1, h_2, h)=0,
\ee
but this does not mean, that $ Z^{a}(K,h_1, h_2)=0$.

In consequence of (40) for the dual lattice we have
 $\widetilde{h}=0$ and
$$
\widetilde{Z}^{\beta}(\widetilde{K},\widetilde{h}_1, \widetilde{h}_2)=
\lim_{\widetilde{h}\to 0} \widetilde{Z}^{(\alpha,\beta)}
(\widetilde{K}, \widetilde{h}_1, \widetilde{h}_2, \widetilde{h})=
\sum_{[\widetilde{\sigma}]}\exp(\widetilde{K}\sum_{\widetilde{r},i}
\wsi{\widetilde{r}}\wsi{\widetilde{r}+i}
$$
\be
+\widetilde{h}_1\sum_{\widetilde{y}=1}^{m}\wsi{n,\widetilde{y}}
\wsi{n,\widetilde{y}+1}+
\widetilde{h}_2\sum_{\widetilde{y}=1}^{m}\wsi{1,\widetilde{y}}
\wsi{1,\widetilde{y}+1}), \quad \beta=a,p
\ee
In result we obtained the partition function of the Ising model
on the cylinder with the free boundary conditions and the defects
on its bases.

Now, taking limits    $h\to\infty$  and  $\widetilde{h}\to 0$  in
(36) and using  (41)-(44), it is not hard to get the duality relations
for the Ising model on the square lattice wrapped on the cylinder
with magnetic fields applied to its bases
 \be
 {\widetilde{Z}^{p}(\widetilde{K},\widetilde{h}_1, \widetilde{h}_2)
 \over\left[\cosh^m 2\widetilde{h}_1\cosh^m 2\widetilde{h}_2
 (\cosh 2\widetilde{K})^{2m(n-3)}
 \right]^{{1/ 2}}}
 =
 { Z^{p}(K, h_1, h_2)+ Z^{p}(K, h_1,-h_2) \over
 \left[\cosh^m 2h_1\cosh^m 2h_2 (\cosh 2K)^{2m(n-3)}
 \right]^{{1/2}}}
 \ee
 \be
 {\widetilde{Z}^{a}(\widetilde{K},\widetilde{h}_1, \widetilde{h}_2)
 \over\left[\cosh^m 2\widetilde{h}_1\cosh^m 2\widetilde{h}_2
 (\cosh 2\widetilde{K})^{2m(n-3)}
 \right]^{{1/ 2}}}
 =
 { Z^{p}(K, h_1, h_2)- Z^{p}(K, h_1,-h_2)
 \over\left[\cosh^m 2h_1\cosh^m 2h_2 (\cosh 2K)^{2m(n-3)}
 \right]^{{1/2}}}.
 \ee
Here we have lattices with sizes $(n-1)\times m$  and $n\times m$
on the right-hand and left-hand sides respectively.

In order to get  the partition function of the 2d Ising model
on the initial lattice with different boundary conditions on the cylinder
bases it is necessary  to consider in (45)-(46) different combinations of limits
 $h_i\to\infty$  and
 ${h_i}\to 0$, $i=1,2$:

 1) the free boundary conditions -- $h_1=h_2=0$,
 which we denote
  $$Z^{\alpha}_{(0,0)}= 2^{-2m}Z^{\alpha}(0,0),\quad\alpha =a,p,$$
 2) the fixed boundary conditions --
  $h_i\to\infty$:
$$ Z^{\alpha}_{(+,+)}=\lim(2\cosh h_1 2\cosh h_1)^{-m} Z^{\alpha}(h_1,h_2),$$
$$\quad Z^{\alpha}_{(+,-)}= \lim(2\cosh h_1 2\cosh h_1)^{-m}
Z^{\alpha}(h_1,-h_2),$$

 3) the mixed boundary conditions  --  $h_1\to\infty$, $h_2=0$
 or $h_2\to\infty$, $h_1=0$:
 $$Z^{\alpha}_{(+,0)}=\lim(4\cosh h_1)^{-m} Z^{\alpha}(h_1,h_2)=
 \lim(4\cosh h_2)^{-m}Z^{\alpha}(h_1,h_2).$$

In consequence of (40) the transition to limits
 $h_i\to\infty$  and
 ${h_i}\to 0$, $i=1,2$
on the initial lattice leads to the following results on the dual lattice:

 1) $\widetilde{h}_1,\widetilde{h}_2\to 0$,
\be
\lim\widetilde{Z}^{\alpha}(\widetilde{K}, \widetilde{h}_1, \widetilde{h}_2) =
(2\cosh\widetilde{K})^{2m} \widetilde{Z}^{\alpha}_{(0,0)}, \quad\alpha=a,p
\ee

 2)  $\widetilde{h}_1\to\infty$,  $\widetilde{h}_2\to\infty$
 \be
 \lim(2\cosh \widetilde{h}_1)^{-m}(2\cosh \widetilde{h}_2)^{-m}
 \widetilde{Z}^{p}( \widetilde{h}_1,
 \widetilde{h}_2)=
 2\left[
 \widetilde{Z}^{p}(\widetilde{K}, \widetilde{K}) +
 \widetilde{Z}^{p}(\widetilde{K}, -\widetilde{K})\right],
 \ee

3) $\widetilde{h}_1\to\infty$,  $\widetilde{h}_2\to 0$
\be
\lim(2\cosh
\widetilde{h}_1)^{-m}\widetilde{Z}^{\alpha}(
\widetilde{h}_1,\widetilde{h}_2) =
(2\cosh \widetilde{K})^{m}2
\widetilde{Z}^{\alpha}(\widetilde{K},0 ),
\ee

Setting in (45), (46) $\widetilde{h}_1=\widetilde{h}_2=\widetilde{K}$
and   respectively ${h}_1=h_2={K}$ we obtain the first two duality relations
for the 2d Ising model with the boundary conditions:
\be
{\widetilde{Z}^{p}_{(0,0)}
 \over (\cosh 2\widetilde{K})^{m(n-3)}}
 =
  {{Z}^{p}(K, {K}) +
 {Z}^{p}({K}, -{K})
 \over (\cosh 2\widetilde{K})^{m(n-3)}}
\ee
\be
{\widetilde{Z}^{a}_{(0,0)}
 \over (\cosh 2\widetilde{K})^{m(n-3)}}
 =
  {{Z}^{p}(K, {K}) -
 {Z}^{p}({K}, -{K})
 \over (\cosh 2\widetilde{K})^{m(n-3)}}
\ee
Recall that here
we have lattices with sizes $(n-1)\times m$  and $n\times m$
on the right-hand and left-hand sides respectively.

Now, using (47)-(49) and taking the corresponding limits in (45), (46),
it is not hard to get the following duality relations:
\be
{(2\cosh\widetilde{K})^{2m} \widetilde{Z}^{p}_{(0,0)}
 \over (\cosh 2\widetilde{K})^{m(n-3)}}
 =
 {Z^{p}_{(+,+)}+ Z^{p}_{(+,-)}
 \over
 (\cosh 2K)^{m(n-3)}},
 \ee
\be
{(2\cosh\widetilde{K})^{2m} \widetilde{Z}^{a}_{(0,0)}
 \over (\cosh 2\widetilde{K})^{m(n-3)}}
 =
 { Z^{p}_{(+,+)}- Z^{p}_{(+,-)}
 \over
 (\cosh 2K)^{m(n-3)}},
 \ee
 \be
 { \widetilde{Z}^{p}(\widetilde{K}, \widetilde{K}) +
 \widetilde{Z}^{p}(\widetilde{K}, -\widetilde{K})
 \over (\cosh 2\widetilde{K})^{m(n-3)}}=
 {Z^{p}_{(0,0)}
 \over
 (\cosh 2K)^{m(n-3)}},
 \ee
 \be
 { \widetilde{Z}^{p}(\widetilde{K}, \widetilde{K}) -
 \widetilde{Z}^{p}(\widetilde{K}, -\widetilde{K})
 \over (\cosh 2\widetilde{K})^{m(n-3)}}=
 {Z^{a}_{(0,0)}
 \over
 (\cosh 2K)^{m(n-3)}},
 \ee
\be
{(2\cosh \widetilde{K})^{m}
\widetilde{Z}^{p}(\widetilde{K},0 )
 \over
 (\cosh 2\widetilde{K})^{m(n-3)}}=
 {Z^{p}_{(0,+)}
 \over
 (\cosh 2K)^{m(n-3)}}.
\ee

Note that in relations (52), (53)
we have lattices with sizes $(n-1)\times m$  and $(n-2)\times m$
on the right-hand and left-hand sides respectively.

Let us make some comments about (53). Using results of exact solution [11]
of the 2d Ising model with magnetic fields applied to the boundaries,
it is not hard to show that
 $ Z^{p}(K, h_1, h_2)- Z^{p}(K, h_1,-h_2)$
is proportional to
$sign(h_1\cdot h_2)$ and therefore at obtaining of (46) the contradiction
after taking limits
$\widetilde{h}_1\to\infty$,
$\widetilde{h}_2\to\infty$ and $\widetilde{h}_1\to\infty$,
$\widetilde{h}_2\to-\infty$   is not appeared.

In the critical point relations (50)-(55) are reduced to
$$ Z^{p}_{(0,0)} =Z^{p}_{(+,+)}+ Z^{p}_{(+,-)}, $$
which was obtained in [12] and it follows from [11].

\bigskip
\centerline{\Large\bf 7. Conclusion}

\medskip

From the duality relation for the nonhomogeneous Ising model one can be obtained
 some usefull concequences.
Using this relation, one can correctly introduce the "mixed" correlation
function of type
$\langle\sigma\mu\sigma'\mu'\rangle$  and  discover their
fermionic content.
In principle  anzats (27) allows to constract the generating functional
depending from external currents
$J(r)$, $\widetilde{J}({r})$ and
$\chi(r)$, where the first two currents are connected with fluctuations
of the order and disorder variables and the last current generates  the
fermionic type exitation.

Undoubtedly the nonhomogeneous duality relation will be usefull for
analysis  of Ising model with random
coupling constants.

However, unfortunaly, we have
not proof of this relation for arbitrary distributions of the coupling
constants and sizes of the lattice. In given paper the duality relation is
proved for the homogeneous case and in the first order for weakly
nonhomogeneous case  (it is not hard to prove one in the second order).  One
can prove this relation for case when the some small numbers of the coupling
constants is chosen arbitrary ones on the background of the rest homogeneous
coupling constants.  The additional argument for correctness of  duality
relation (27) is the direct check  of one on the small lattices with sizes
$m,n=2,3$.

\medskip
We would like thank A.~Belavin, Vl.~Dotsenko, M.~Lashkevich
for usefull remarks connected with subject of this paper.

V.S. thanks Dr. A.~Morozov for the hospitality and the exellent
conditions at ITEP, where this paper has been finished.

\newpage
\centerline{\bf REFERENCES}

\medskip

\begin{enumerate}
\item H.A.~Kramers, G.H.~Wannier, Phys. Rev.
{\bf 60}, 252 (1941).
\item L.~Onsager, Phys. Rev. {\bf 65}, 117
(1944).
\item L.P.~Kadanoff, H.~Ceva, Phys. Rev. {\bf B3}, 3918
(1971).
\item C.A.~Hurst, H.S.~Green, J. Chem. Phys. {\bf 33}, 1059
(1960)
\item F..~Berezin, Usp. Mat. Nayk {\bf 24}, 3 (1969).
\item A.I.~Bugrij, {\it "Partition Function of Planar Ising Model
on Finite Size Lattice" }, Preprint ITP-85-114R (1985).
\item A.I.~Bugrij, V.N.~Shadura, Teor. Mat. Fyz. {\bf
 103}, 388 (1995).
\item Vik..~Dotsenko, Usp. Fyz. Nayk {\bf 165}, 482
(1995).
\item P.~Di Francesco, H.~Saleur, J.B.~Zuber, Preprint SPh T/87-097,
Saclay (1987).
\item A.~Belavin, A.~Polyakov, A.~Zamolodchikov, Nuc. Phys.
{\bf B241}, 333 (1984).
\item  A.I.~Bugrij, V.N.~Shadura, Phys. Lett. A, {\bf 150A}, n.3,4, 171 (1990)
\item  J.L.Cardy, {\it "Conformal Invariance and Statistical Mechanics"},
in "Fields, Strings and Critical Phenomena" ed. by E.~Brezin and
J.~Zinn-Justin, Elsevier Science Pub. (1989)
\end{enumerate}
\end{document}